\documentclass[english,aps,prl,reprint,twocolumn]{revtex4-1}
\usepackage{prettyref}
\usepackage{amsmath}
\usepackage{amssymb}
\usepackage{graphicx}

\newcommand{\T}{\mathbb{T}}

\begin{document}

\title{Optimal wall-to-wall transport by incompressible flows}

\author{Ian Tobasco$^1$ and Charles R.~Doering$^{1,2,3}$}

\affiliation{$^1$Department of Mathematics, University of Michigan, Ann Arbor, MI 48109-1043}
\affiliation{$^2$Department of Physics, University of Michigan, Ann Arbor, MI 48109-1040}
\affiliation{$^3$Center for the Study of Complex Systems, University of Michigan, Ann Arbor, MI 48109-1107}

\date{\today}
\begin{abstract}
We consider wall-to-wall transport of a passive tracer by divergence-free velocity vector fields~$\mathbf{u}$.
Given an enstrophy budget $\langle |\nabla \mathbf{u}|^{2} \rangle \le Pe^{2}$ we construct steady two-dimensional flows that transport at rates $Nu(\mathbf{u}) \gtrsim Pe^{2/3}/(\log Pe)^{4/3}$ in the large enstrophy limit. 
Combined with the known upper bound $Nu(\mathbf{u})\lesssim Pe^{2/3}$ for any such enstrophy-constrained flow, we conclude that maximally transporting flows satisfy $Nu\sim Pe^{2/3}$ up to possible logarithmic corrections.
Combined with known transport bounds in the context of Rayleigh-B\'enard convection this establishes that while suitable flows approaching the ``ultimate'' heat transport scaling $Nu\sim Ra^{1/2}$ exist, they are not always realizable as buoyancy-driven flows.
The result is obtained by exploiting a connection between the wall-to-wall optimal transport problem and a closely related class of singularly perturbed variational problems arising in the study of energy-driven pattern formation in materials science.
\end{abstract}

\maketitle

\noindent
{\it Introduction --}  Modeling, measuring, and controlling the transport properties of incompressible flows is a fundamental aspect of fluid mechanics with myriad applications in engineering and the applied sciences.
In some cases the transport of heat or trace concentrations of impurities is {\it passive}, i.e., the thermal energy or mass markers are carried without essentially altering the flow. 
In other settings the transport is {\it active} as is the situation when heat or dissolved mass alters the fluid density to produce buoyancy forces in the presence of a gravitational field, or more generally for momentum transport responsible for the transmission of drag forces.
In this Letter we study the primary problem of passive tracer transport between parallel walls by a combination of molecular diffusion and fluid advection when the tracer concentration is set at the walls to determine the maximum transport increase over diffusion alone that incompressible flows of a given intensity can induce.
The results are of interest in their own right but they also have implications for the active transport problem of buoyancy-driven turbulent convection.

The mathematical formulation is as follows.
The spatial domain $\Omega$ is periodic in $x$ and $y$ with rigid walls at $z=0$ and $z=1$.
The tracer field $T(x,y,z,t)$, referred to as temperature, satisfies the advection-diffusion equation
\begin{equation}
\partial_{t}T+\mathbf{u}\cdot\nabla T=\Delta T\label{eq:advectiondiffusion}
\end{equation}
in $\Omega$ with boundary conditions $T|_{z=0}=1$ and $T|_{z=1}=0$ where $\mathbf{u}=\hat{\mathbf{i}} u + \hat{\mathbf{j}} v + \hat{\mathbf{k}} w$ is an {arbitrary} divergence-free velocity field with no-slip boundary conditions $\mathbf{u}|_{\partial\Omega}=0$.
These are dimensionless variables: lengths are measured in units of $h$, time in units of $h^2/\kappa$, and $\mathbf{u}$ in units of $\kappa/h$ where $h$ is the wall-to-wall distance and $\kappa$ is the thermal diffusivity.
$T$ is measured in units of the temperature drop across the layer.

The Nusselt number $Nu$ is a measure of enhancement of wall-to-wall transport relative to pure conduction: it is the ratio of total convective to conductive vertical heat flux given here by
\begin{equation}
Nu(\mathbf{u})
=1+\left\langle w T\right\rangle
\end{equation}
where $\left\langle \cdot\right\rangle $ indicates the long-time and space average.
We are concerned with the design of incompressible flows that, subject to an intensity budget $\langle |\nabla\times \mathbf{u}|^{2} \rangle = \langle |\nabla\mathbf{u}|^{2} \rangle \le Pe^2$, maximize wall-to-wall heat transport:
\begin{equation}
F(Pe) \ =\max_{\left\langle |\nabla\mathbf{u}|^{2}\right\rangle \leq Pe^{2}}\,Nu(\mathbf{u}).\label{eq:walltowalldefn}
\end{equation}
The non-dimensional P\'eclet number $Pe$ is a measure of advective intensity relative to that of diffusion and we take it to be the (maximum allowable) root mean square rate of strain, equivalent here to the square root of the mean enstrophy. We are particularly interested in the behavior of the maximal transport $F(Pe)$ as $Pe\to\infty$.

Our motivation is twofold.
First, while the wall-to-wall optimal transport problem is both easy to state and natural from a practical point of view---the power required to sustain such a Newtonian fluid flow is proportional to its mean square rate of strain---it turns out to be quite challenging to identify the salient properties of optimal flows in the large enstrophy limit.
In the energy-constrained problem where the budget is set by the kinetic energy, the optimal transport scaling is captured by a simple convection roll design \citep{hassanzadeh2014wall}.
The enstrophy-constrained problem considered here is substantially more subtle: numerical work \citep{hassanzadeh2014wall,SD} suggests that optimal flows are not simple convection rolls, but instead more complex designs featuring near wall recirculation zones whose fine-scale features {are yet to be described}. 

Second, the wall-to-wall optimal transport problem can be used to derive absolute limits on the rate of heat transport in Rayleigh-B\'enard convection (RBC), the buoyancy-driven flow of fluid heated from below and cooled from above \citep{Rayleigh1916}.
In the Boussinesq approximation RBC is modeled by supplementing \prettyref{eq:advectiondiffusion} with the forced Navier-Stokes equations
\begin{equation}
\partial_{t}\mathbf{u}+\mathbf{u}\cdot\nabla\mathbf{u}+\nabla p = Pr \Delta\mathbf{u} + Pr\, Ra \, \hat{\mathbf{k}} \, T\label{eq:NSE}
\end{equation}
for the divergence-free velocity field $\mathbf{u}(x,y,z,t)$ where $Pr$ and $Ra$ are the Prandtl and Rayleigh numbers.
It is a long-standing question to determine rigorous $Nu$\textendash $Pr$\textendash $Ra$ relationships for RBC.
The best known rigorous result that applies uniformly in $Pr$ for no-slip boundaries is $Nu \lesssim Ra^{1/2}$ for $Ra\gg1$  \citep{howard1963heat,busse1969howards,doering1996variational,seis2015}, i.e., the so-called ``ultimate'' heat transport scaling \cite{spiegel1971}.

Dotting $\mathbf{u}$ into equation \prettyref{eq:NSE}, integrating by parts and time averaging reveals that $\left\langle |\nabla\mathbf{u}|^{2}\right\rangle =Ra\cdot(Nu-1)$.
Thus, by the definition \prettyref{eq:walltowalldefn} of wall-to-wall optimal transport,
\[
Nu\leq F\left( Ra\cdot(Nu-1) \right).
\]

This optimal wall-to-wall approach for proving absolute limits on the rate of heat transport by RBC flows was proposed as a potentially more powerful alternative to the established methods \citep{hassanzadeh2014wall}.
Here the advection-diffusion equation \prettyref{eq:advectiondiffusion} is imposed as a point-wise constraint, whereas previous analyses utilized only certain mean/moment balances derived from the governing equations.
Therefore, the wall-to-wall optimal transport approach has the propensity to produce better bounds on $Nu$ as a function of $Ra$.
Moreover, it produces explicit incompressible flow fields realizing optimal transport which are of interest in their own right.

The aforementioned methods for deriving upper bounds in RBC applied here prove that $F(Pe) \lesssim Pe^{2/3}$
for $Pe \gg 1$ (see, e.g., \citep{SD}). In this Letter we explore the sharpness of this \emph{a priori} estimate insofar as its scaling is concerned.
Our methods shed light on the nature of maximally transporting flows and make precise what is gained in the context of rigorous bounds in RBC by enforcing \prettyref{eq:advectiondiffusion} pointwise.
To this end we construct steady no-slip incompressible flows $\left\{\mathbf{u}_{Pe}\right\}$ such that
\begin{equation}
\left\langle |\nabla \mathbf{u}_{Pe}|^{2}\right\rangle \leq Pe^{2}\quad\text{and}\quad Nu(\mathbf{u}_{Pe}) \gtrsim \frac{Pe^{2/3}}{(\log Pe)^{4/3}}\label{eq:mainclaim}
\end{equation}
for all $Pe \gg 1$ to conclude that incompressible flows can indeed achieve $Nu\sim Pe^{2/3}$ up to possible logarithmic corrections. 
To obtain the result we exploit an interesting and perhaps unexpected connection between the wall-to-wall optimal transport problem and optimal design problems arising for energy-driven pattern formation in materials science \citep{kohn2007energy}.

The rest of this Letter is organized as follows.
First we derive a variational formulation for the transport rate of an arbitrary steady incompressible flow.
Then we introduce a Lagrange multiplier for the enstrophy constraint to discover a direct analog of Howard's variational problem for RBC \citep{howard1963heat} in the context of wall-to-wall optimal transport.
The resulting problem is reminiscent of questions in materials science, inspiring construction of the nearly optimal flows.
We end with further discussion of connections between fluid dynamical and materials science variational problems.

\medskip
\noindent
{\it Variational formulation for transport rates --}
We begin by deriving variational formulations for the rate of heat transport, inspired by variational formulations for the effective diffusivity in periodic homogenization \citep{fannjiang1994convection}. (See also \cite{avellaneda1991integral,milton1990characterizing}.)
The methods laid out there for periodic domains can be adapted to our domain as well.
And we may restrict attention to steady velocity fields: indeed, the maximal unsteady transport rate is no less than its steady counterpart.

The steady temperature deviation $\theta=T+z-1$ satisfies
\begin{equation}
\mathbf{u}\cdot\nabla\theta=\Delta\theta+w \label{eq:deviationeqn}
\end{equation}
with boundary conditions $\theta|_{\partial\Omega}=0$.
Then $Nu(\mathbf{u})-1=\left\langle |\nabla\theta|^{2}\right\rangle =\left\langle w \theta\right\rangle$ and we can state dual variational formulations for it:
\begin{align}
&Nu(\mathbf{u}) -1  \nonumber \\
& \ \ = \min_{\eta:\eta|_{\partial\Omega}=0}  \left\langle |\nabla\eta|^{2}\right\rangle + \left\langle |\nabla\Delta^{-1} (-w+\mathbf{u}\cdot\nabla\eta)|^{2}\right\rangle  \label{eq:primalNu} \\
& \ \ =\max_{\xi:\xi|_{\partial\Omega}=0} 2 \left\langle  w \xi\right\rangle -\left\langle |\nabla\Delta^{-1}\mathbf{u}\cdot\nabla\xi|^{2}\right\rangle -\left\langle |\nabla\xi|^{2}\right\rangle \label{eq:dualNu}
\end{align}
where $\Delta^{-1}$ is the inverse Laplacian operator with Dirichlet boundary conditions on $\partial \Omega$.

To see these consider the pair of equations
\[
\pm\mathbf{u}\cdot\nabla\theta_{\pm}=\Delta\theta_{\pm}+w.
\]
Then  $\xi=\frac{1}{2}(\theta_{+}+\theta_{-})$
and $\eta=\frac{1}{2}(\theta_{+}-\theta_{-})$ satisfy
\begin{align}
\mathbf{u}\cdot\nabla\eta & =\Delta\xi+ w ,\label{eq:systema}\\
\mathbf{u}\cdot\nabla\xi & =\Delta\eta\label{eq:systemb}
\end{align}
and either variable can be eliminated to produce 
\begin{align}
\mathbf{u}\cdot\nabla\Delta^{-1}\mathbf{u}\cdot\nabla\eta & =\Delta\eta+\mathbf{u}\cdot\nabla\Delta^{-1} w,\\
\mathbf{u}\cdot\nabla\Delta^{-1}\mathbf{u}\cdot\nabla\xi & =\Delta\xi+ w. \label{eq:xieqn}
\end{align}
These are the Euler-Lagrange equations for the well-posed problems
\prettyref{eq:primalNu} and \prettyref{eq:dualNu} so it remains only
to verify that the {optimal $\eta$ and $\xi$ appearing there achieve} the desired
value of $Nu(\mathbf{u})-1$. 

First consider the optimal $\eta$.
Testing \prettyref{eq:systemb} against $\xi$ and integrating by parts shows that $\nabla\xi\perp\nabla\eta$ in $L^{2}(\Omega)$.
Hence,
\[
Nu(\mathbf{u})-1=\left\langle |\nabla\theta_{+}|^{2}\right\rangle =\left\langle |\nabla\xi|^{2}\right\rangle +\left\langle |\nabla\eta|^{2}\right\rangle 
\]
and as $\xi$ is recovered from $\eta$ through \prettyref{eq:systema}
this verifies \prettyref{eq:primalNu}. 

Next consider the optimal $\xi$. A similar
integration by parts argument involving \prettyref{eq:systema} and
\prettyref{eq:xieqn} shows that $w \perp\eta$ in $L^{2}(\Omega)$
and that
\begin{equation}
\left\langle w\xi\right\rangle =\left\langle |\nabla\Delta^{-1}\mathbf{u}\cdot\nabla\xi|^{2}\right\rangle +\left\langle |\nabla\xi|^{2}\right\rangle .\label{eq:energybalance}
\end{equation}
Therefore,
\[
Nu(\mathbf{u})-1=\left\langle w \theta_{+}\right\rangle =\left\langle w \xi\right\rangle 
\]
and combining this with \prettyref{eq:energybalance} gives \prettyref{eq:dualNu}.

The change of variables $(\theta_{+},\theta_{-})\leftrightarrow(\eta,\xi)$ is key to these formulations.
It was also used in the case of energy-constrained wall-to-wall optimal transport \citep{hassanzadeh2014wall} where it was observed that $\eta$ depends only on $z$ permitting asymptotic solution of the Euler-Lagrange equations.
Such simplification does not occur in the enstrophy-constrained case but we can still exploit \prettyref{eq:dualNu} to deduce rigorous lower bounds.

\smallskip
\noindent
{\it Nearly optimal velocity fields --}
Introduce a Lagrange multiplier for the enstrophy constraint and consider
\[
M(\lambda)=\max_{\mathbf{u}} \left\{Nu(\mathbf{u})-\lambda^{2}\left\langle |\nabla\mathbf{u}|^{2}\right\rangle \right\}
\]
for $\lambda\ll1$. 
Then \prettyref{eq:dualNu} and straightforward rescalings imply
\[
M(\lambda)-1=\max_{a}\left\{2a-a^{2}\cdot\min_{\langle w \xi \rangle =1}E_{\frac{\lambda}{a}}(\mathbf{u},\xi)\right\}
\]
where 
\begin{equation}
E_{\epsilon}(\mathbf{u},\xi)=\left\langle |\nabla\Delta^{-1}\mathbf{u}\cdot\nabla\xi|^{2}\right\rangle +\epsilon\left\langle |\nabla\mathbf{u}|^{2}+|\nabla\xi|^{2}\right\rangle .\label{eq:energy}
\end{equation}

This form of the problem, $\min E_\epsilon$, bears an interesting resemblance both to Howard's variational problem for RBC bounds \citep{howard1963heat} and also to problems originally arising in the study of energy-driven pattern formation in materials science (more on this later).
For now we assert that 
\[
\epsilon^{1/2}\lesssim\min_{\langle w \xi \rangle =1}E_{\epsilon}(\mathbf{u}, \xi )\lesssim\epsilon^{1/2}\log\frac{1}{\epsilon}
\]
for $\epsilon\ll1$. The lower bound is the direct translation of the known upper bound $F(Pe) \lesssim Pe^{2/3}$ to this minimization problem in the case of steady velocities.
Our focus is on the upper bound: next we construct test fields $(\mathbf{u}_{\epsilon},\xi_{\epsilon})$ satisfying the net flux constraint $\langle w_{\epsilon} \xi_{\epsilon} \rangle =1$ such that 
\begin{equation}
\left\langle |\nabla\mathbf{u}_{\epsilon}|^{2}\right\rangle \sim\epsilon^{-1/2}\log\frac{1}{\epsilon}\ \text{ and }\ E_{\epsilon}(\mathbf{u}_{\epsilon},\xi_{\epsilon})\lesssim\epsilon^{1/2}\log\frac{1}{\epsilon}\label{eq:upperboundonE}
\end{equation}
for $\epsilon\ll1$.
After performing the construction we will undo the rescalings to recover the main result \prettyref{eq:mainclaim}.

\begin{figure}
\center\includegraphics[width=\columnwidth]{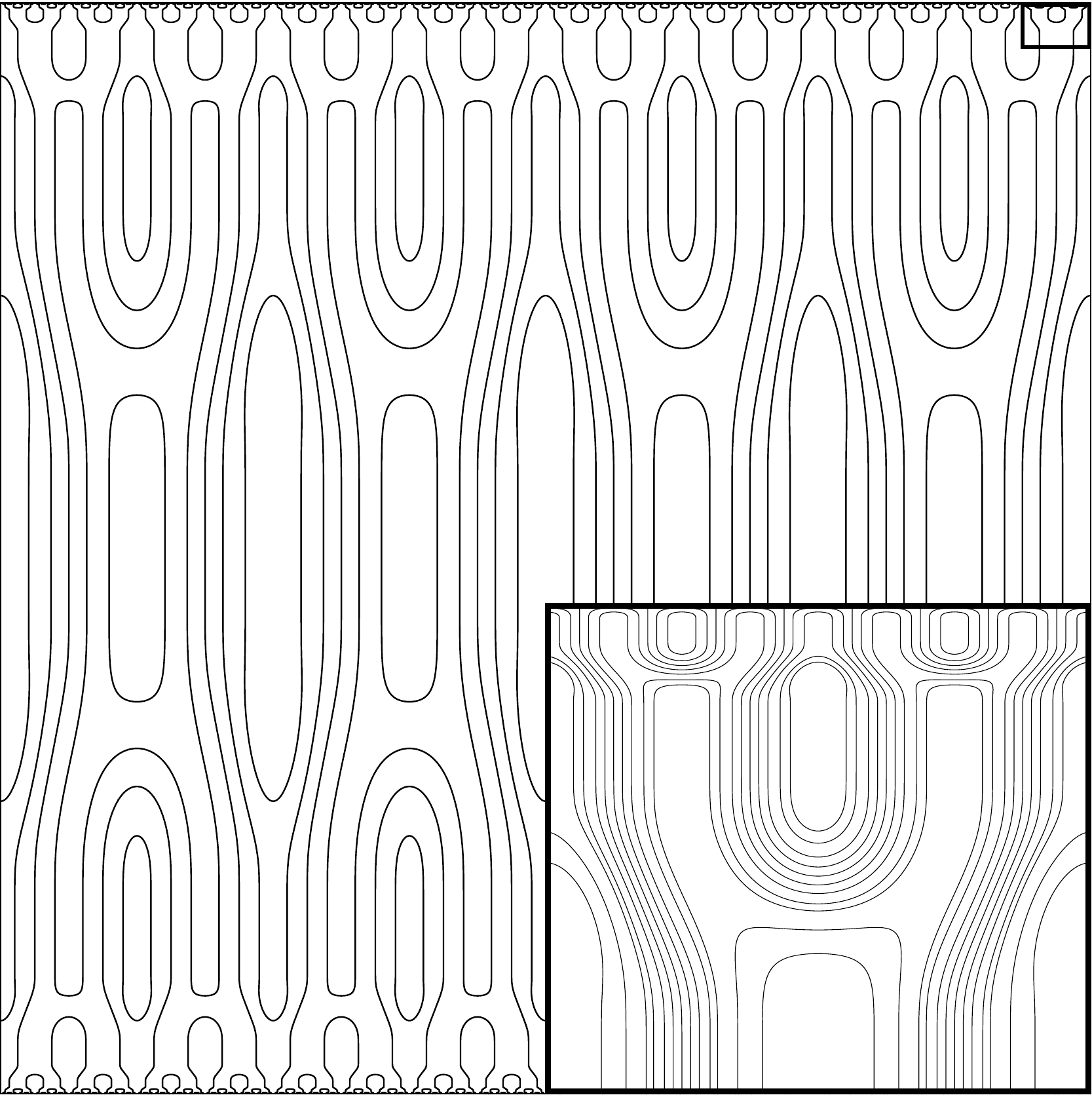}

\caption{Schematic streamlines of the nearly optimal flow. Streamlines branch and self-similarly refine from  bulk to boundary layer; this terminates once the design resembles isotropic convection rolls. Inset shows structure at the wall.}
\label{fig:coarseningconstruction}
\end{figure}

\medskip
\noindent
{\it The branching construction --}
A judiciously chosen streamfunction $\psi(x,z)$ describes a two-dimensional (2D) divergence-free velocity field {$\mathbf{u}=(-\partial_{z}\psi,0,\partial_{x}\psi)$} that is well-aligned wall-to-wall and whose direction fluctuates at a length-scale $\ell(z)$ depending monotonically on the distance to the wall.
Choose $n$ points $\{z_{k}\}_{k=1}^{n}$ satisfying $\frac{1}{2} < z_{1} < z_{2} \cdots < z_{n} <1$ and let $l_{k}=\ell(z_{k})$ be the length-scale at the $k$th cross section with $\psi(x,z_{k})=\psi_{k}(x)=c_{0}\sqrt{2}l_{k}\cos(2\pi l_{k}^{-1} x)$.
(The $l_{k}$s will be compatible with $2\pi$-periodicity and the constant $c_{0}$ will be chosen below.)
For $1\leq k\leq n-1$ extend the streamfunction across the $k$th transition layer $\Omega_{k}=\T_{x}\times[z_{k},z_{k+1}]$ ($\T_{x}$ is the periodic $x$-interval) by
\[
\psi(x,z)=f\left(\frac{z-z_{k}}{z_{k+1}-z_{k}}\right)\psi_{k}(x)+f\left(\frac{z_{k+1}-z}{z_{k+1}-z_{k}}\right)\psi_{k+1}(x)
\]
where $f\in C^{\infty}([0,1])$ is a cutoff function, fixed once and
for all. We require the Pythagorean condition 
\[
(f(t))^{2}+(f(1-t))^{2}=1
\]
and also that $f(0)=1$, $f(1)=0$, and $f'(0)=f'(1)=0$.
We let $\psi(x,z)=\psi_{1}(x)$ in the bulk domain $\Omega_{bulk}=\T_{x}\times[\frac{1}{2},z_{1}]$, $\psi(x,z)=f(\frac{z-z_{n}}{1-z_{n}})\psi_{n}(x)$ in the thermal boundary layer $\Omega_{bl}=\T_{x}\times[z_{n},1]$, and extend it by even reflection across $z=1/2$ to all of $\Omega$.
See Figure \ref{fig:coarseningconstruction} above.

Next we choose the test field $\xi$.
The wall-to-wall velocity component $w$ and $\xi$ must be well-correlated to enforce the net flux constraint $\langle w \xi \rangle =1$ so we fix $\xi=w$.
Then, by the $L^{2}$-orthonormality of $\{c_0^{-1}\psi_{k}'\}$, 
\begin{align*}
\frac{1}{2c_{0}^{2}} \langle w^{2} \rangle  & = \left(\int_{\frac{1}{2}}^{z_{1}}+\sum_{k=1}^{n-1}\int_{z_{k}}^{z_{k+1}}+\int_{z_{n}}^{1}\right)||c_{0}^{-1}\partial_{x}\psi||_{L_{x}^{2}}^{2} \, dz\\
 & =z_{n}-\frac{1}{2}+(1-z_{n})\int_{0}^{1}f^{2}=\frac{1}{2}z_{n}.
\end{align*}
Choosing $c_{0}=z_{n}^{-1/2}$ satisfies the flux constraint.

We proceed to bound the terms appearing in $E_{\epsilon}$ in \prettyref{eq:energy}.
Let $\delta_{k}=|z_{k+1}-z_{k}|$ be the thickness of the $k$th transition layer $\Omega_{k}$, let $\delta_{bl}=|1-z_{n}|$ be the thickness of the thermal boundary layer $\Omega_{bl}$, and let $\delta_{bulk}=|z_{1}-\frac{1}{2}|$ be the thickness of the bulk domain $\Omega_{bulk}$.
Recall that $l_{k}=\ell(z_{k})$ is the horizontal length-scale at the $k$th cross section $\T_{x}\times\{z_{k}\}$, and let $l_{bulk}=l_{1}$ and $l_{bl}=l_{n}$ be the horizontal length-scales appearing in $\Omega_{bulk}$ and $\Omega_{bl}$ respectively. Similarly define $z_{bulk} = z_1$ and $z_{bl} = z_n$.
We then have the following estimates for the advection and enstrophy terms:
\begin{align}
\left\langle |\nabla\Delta^{-1}\mathbf{u}\cdot\nabla w |^{2}\right\rangle  & \lesssim\int_{z_{bulk}}^{z_{bl}}(\ell')^{2}dz+l_{bl},\label{eq:advectionestimate}\\
\left\langle |\nabla\mathbf{u}|^{2}\right\rangle  & \sim\frac{1}{l_{bulk}^{2}}+\int_{z_{bulk}}^{z_{bl}}\frac{1}{\ell^{2}}dz+\frac{1}{l_{bl}}.\label{eq:enstrophyestimate}
\end{align}
Note for these to hold we must take $\delta_{bulk}\sim1$, $l_{k}\lesssim\delta_{k}$ and $l_{bl}\sim\delta_{bl}$, and finally $l_{k+1}\sim l_{k}$ and $|l_{k+1}-l_{k}|\sim l_{k}$ for all $k$.
Under these restrictions we conclude that 
\[
E_{\epsilon} \lesssim \epsilon \, \frac{1}{l_{bulk}^{2}}+\int_{z_{bulk}}^{z_{bl}} \left[ (\ell')^{2}+\epsilon\frac{1}{\ell^{2}} \right] dz + l_{bl} + \epsilon\frac{1}{l_{bl}}
\]
with a constant that only depends on those implicit in the assumptions.

Consider minimizing the righthand side above over all $\ell(z)$.
The optimal $\ell$ satisfies $\ell'=\epsilon^{1/2}\ell^{-1}$ on $(z_{bulk},z_{bl})$.
It is natural to think of solving this equation on $(\frac{1}{2},1)$ with the initial condition $\ell(1)=0$ leading immediately to the power law
\[
\ell(z)\sim\epsilon^{1/4}(1-z)^{1/2}.
\]
Choosing $\ell_{bulk}\sim\epsilon^{1/4}$ and $\ell_{bl}\sim\epsilon^{1/2}$ we are led by \prettyref{eq:advectionestimate} and \prettyref{eq:enstrophyestimate} to the estimates $E_{\epsilon}\lesssim\epsilon^{1/2}\log\frac{1}{\epsilon}$ and $\left\langle |\nabla\mathbf{u}_{\epsilon}|^{2}\right\rangle \sim\epsilon^{-1/2}\log\frac{1}{\epsilon}$ for $\epsilon \ll 1$.

Now we prove \prettyref{eq:upperboundonE}.
Take $\ell(z)=2^{-n}(1-z)^{1/2}$ and fix the interpolation points $z_{k}=1-2^{-2k}$ so that $\delta_{k}=\frac{3}{4}\cdot 2^{-2k}$ and $l_{k}=2^{-k-n}$.
Given $\epsilon>0$, let $n$ satisfy $\frac{1}{4}\log_{2}\frac{1}{\epsilon}\leq n<\frac{1}{4}\log_{2}\frac{1}{\epsilon}+1$
and note that $\epsilon\sim2^{-4n}$.
Since $\delta_{bulk}\sim1$, $\delta_{k}\sim2^{-2k}$ and $l_{k}\sim 2^{-k-n}$, and $l_{k}=2l_{k+1}$ we see that the requirements for \prettyref{eq:advectionestimate} and \prettyref{eq:enstrophyestimate} hold.
Therefore, the arguments above prove the validity of \prettyref{eq:upperboundonE}.

\medskip
\noindent
{\it Rescalings and the Lagrange multiplier --}
We can now deduce our main result \prettyref{eq:mainclaim}.
Let $(\mathbf{u}_{\epsilon},\xi_{\epsilon})$ be as in \prettyref{eq:upperboundonE}.
Let $\epsilon=\lambda/a$ where $a,\lambda>0$ are to be chosen, and perform the rescalings $\tilde{\mathbf{u}}=a^{1/2}\lambda^{-1/2}\mathbf{u}_{\lambda/a}$ and $\tilde{\xi}=a^{1/2}\lambda^{1/2}\xi_{\lambda/a}$.
Then, according to \prettyref{eq:upperboundonE}, 
\[
c_{1}\frac{a^{3/2}}{\lambda^{3/2}}\log\frac{a}{\lambda}\leq\left\langle |\nabla\tilde{\mathbf{u}}|^{2}\right\rangle \leq c_{2}\frac{a^{3/2}}{\lambda^{3/2}}\log\frac{a}{\lambda}
\]
and 
\[
Nu(\tilde{\mathbf{u}})\geq2a-ca^{3/2}\lambda^{1/2}\log\frac{a}{\lambda}+\lambda^{2}\left\langle |\nabla\tilde{\mathbf{u}}|^{2}\right\rangle 
\]
where $c_{1}$, $c_{2}$, and $c$ are independent of all parameters.

We maximize in $a$. The optimal $a$ satisfies a transcendental equation so to capture the asymptotics we set $a=\frac{\theta_1}{\lambda\log^{2}\lambda}$ where $\theta_1$ depends only on $c_{1}$ and $c$.
Then for $\lambda\ll1$, $\tilde{\mathbf{u}}$ satisfies
\[
\left\langle |\nabla\tilde{\mathbf{u}}|^{2}\right\rangle \leq2c_{2}\frac{\theta_1^{3/2}}{\lambda^{3}\log^{2}\lambda}\quad\text{and}\quad Nu(\tilde{\mathbf{u}})\gtrsim\frac{1}{\lambda\log^{2}\lambda}.
\]

Finally, we can prove \prettyref{eq:mainclaim}.
We do so by choosing the Lagrange multiplier to satisfy $\lambda=\theta_2 Pe^{-2/3}(\log Pe)^{-2/3}$ where $\theta_2$ depends only on $c_1$, $c_2$, and $c$.
Then \prettyref{eq:mainclaim} follows from the rescalings performed above.

Observe that $\epsilon\sim Pe^{-4/3}(\log Pe)^{2/3}$.
Thus, in terms of the original parameters, our nearly optimal velocity fields $\{\mathbf{u}_{Pe}\}$ exhibit horizontal fluctuations at a length-scale
\[
\ell(z)\sim Pe^{-1/3}(\log Pe)^{1/6} \, (1-z)^{1/2}
\]
{for $z\in(z_{bulk},z_{bl})$. In the bulk the horizontal length-scale} obeys $l_{bulk}\sim Pe^{-1/3}(\log Pe)^{1/6}$, while in the thermal boundary layer  $l_{bl}\sim Pe^{-2/3}(\log Pe)^{1/3}$. 

\medskip
\noindent
{\it Discussion --}  The ultimate result of this Letter is that there exist incompressible flows satisfying suitable boundary conditions and intensity constraints that transport heat by \prettyref{eq:advectiondiffusion} and saturate, modulo logarithmic corrections, the upper bound $Nu \lesssim Ra^{1/2}$ that holds for any RBC flow.
It does {\it not}, however, establish the existence of solutions to the {\it full} Boussinesq system  \prettyref{eq:advectiondiffusion} and \prettyref{eq:NSE} that realize such transport.
The actual behavior of large Rayleigh number RBC transport remains an open question mathematically.
We note here, however, the recent result obtained in \cite{whitehead2011ultimate} for RBC transport between stress-free boundaries in 2D that states that $Nu\lesssim Ra^{5/12}$ uniformly in $Pr$.
Combining this bound with the results of this Letter, and the fact that the optimal transport between stress-free boundaries is no smaller than between no-slip boundaries \cite{SD}, we conclude that buoyancy forces {\it cannot} achieve---or even approach---the actual optimal wall-to-wall transport in 2D stress-free RBC .

Mathematical analysis of upper bounds on the rate of heat transport in RBC goes back at least to Howard \citep{howard1963heat} who, employing suitable mean/moment balance laws, introduced the variational problem
\begin{equation}
m(\lambda)=\min_{\langle w \xi \rangle =1}\left\langle |\overline{w\xi}-1|^{2}\right\rangle +\lambda\left\langle |\nabla\mathbf{u}|^{2}\right\rangle\cdot\left\langle |\nabla\xi|^{2}\right\rangle \label{eq:Howardpblm1}
\end{equation}
where $\overline{f}$ stands for the average in the periodic variables $x$ and $y$. 
Here we introduce the related problem
\begin{equation}
\tilde{m}(\epsilon)=\min_{\langle w \xi \rangle =1}\left\langle |\overline{w\xi}-1|^{2}\right\rangle +\epsilon\left\langle |\nabla\mathbf{u}|^{2}+|\nabla\xi|^{2}\right\rangle\label{eq:Howardpblm2}
\end{equation}
and note that $m(\lambda)\sim\lambda^{1/3}$ for $\lambda\ll1$, while $\tilde{m}(\epsilon)\sim\epsilon^{1/2}$ for $\epsilon\ll1$.
The former was obtained by Howard and Busse in their groundbreaking works \citep{howard1963heat,busse1969howards}. The lower bounds implicit in both of these scalings are equivalent to the upper bound $Nu\lesssim Ra^{1/2}$.

Our interest in \prettyref{eq:Howardpblm1} and \prettyref{eq:Howardpblm2} is in their relation to wall-to-wall optimal transport. 
We showed above that the steady wall-to-wall problem is equivalent to the minimization of $E_{\epsilon}(\mathbf{u},\xi)$ under a net flux constraint $\langle w \xi \rangle =1$ (see equation \eqref{eq:energy} and the surrounding discussion).
Now we decompose the advection term in $E_{\epsilon}$ as
\[
\left\langle |\nabla\Delta^{-1}\text{div}\,\mathbf{u}\xi|^{2}\right\rangle =\left\langle |\overline{w\xi}-1|^{2}\right\rangle +\mathcal{Q}(\mathbf{u}\xi)
\]
where $\mathcal{Q}$ is the positive semi-definite quadratic form 
\[
\mathcal{Q}(\mathbf{m})=\min_{\substack{\mathbf{w}:\text{div}\,\mathbf{w}=0}
}\left\langle |\mathbf{w}+\mathbf{m}-\overline{\mathbf{m} \cdot \hat{\mathbf{k}}} \ \hat{\mathbf{k}}|^{2}\right\rangle .
\]
Evidently this new term $\mathcal{Q}$, not present in \prettyref{eq:Howardpblm1} and \prettyref{eq:Howardpblm2}, arises from the advection-diffusion constraint \prettyref{eq:advectiondiffusion}. 

As shown in this Letter, the wall-to-wall optimal transport approach cannot result in a significantly improved upper bound on heat transport in turbulent RBC, i.e., improvement cannot come in the form $Nu\lesssim Ra^{\alpha}$ with $\alpha<\frac{1}{2}$.
Still, the quadratic form $\mathcal{Q}$ does play a non-trivial role in our construction of nearly optimal flows: it is precisely this form that supplies the term $\int_{z_{bulk}}^{z_{bl}}(\ell')^{2}dz$ in the advection estimate \prettyref{eq:advectionestimate}.
So, at the level of constructions, $\mathcal{Q}$ is what gives rise to the logarithmic correction 
in  \prettyref{eq:mainclaim}.
It remains to be seen if it actually modifies the behavior of {the optimal transport} function $F(Pe)$.

The branching flow structure described in this Letter is similar to Busse's ``multi $\alpha$'' technique \citep{busse1969howards} for the analysis of Howard's problem.
Busse observed that \prettyref{eq:Howardpblm1} cannot be solved as $\lambda \rightarrow 0$ by flows featuring only one horizontal mode.
Instead, increasingly more horizontal modes emerge as $\lambda \to 0$  with wavenumbers $\{\alpha_{k}\}_{k=1}^{n}$ depending on the distance to the wall.
The resulting picture is similar to that presented here albeit with significantly different vertical and horizontal length-scales $\{\delta_{k}\}_{k=1}^{n}$ and $\{l_{k}\}_{k=1}^{n}$.

But Busse's work was {\it not} how we came upon the idea for this sort of flow in wall-to-wall optimal transport.
Instead we observed that the functional $E_\epsilon$ in \prettyref{eq:energy} shares striking similarities with various functionals arising in the study of energy-driven pattern formation in materials science \citep{kohn2007energy} where emergent multiple-scale structures are commonly referred to as ``branching''.
Three examples come to mind: domain branching in uniaxial ferromagnetics \citep{choksi1998bounds,choksi1999domain}, branching of twins near an austenite\textendash twinned-martensite interface \citep{kohn1992branching,kohn1994surface}, and self-similar blistering patterns in a biaxially compressed thin
elastic film \citep{ortiz1994morphology,jin2001energy,belgacem2000rigorous}.
The morphology of low energy states in these examples results from the competition between a non-convex lowest order term (e.g., in micromagnetics, the anisotropy and magnetostatic energies) and a higher order convex regularization (e.g., the exchange energy).
Branching efficiently matches boundary conditions to low-energy states in the bulk.
Continuing with the analogy of micromagnetics, Privorotski\u{\i}'s construction is to our branching flow construction what the Landau-Lifshitz structure is to single mode convection rolls.
Regarding elastic blistering, we see a parallel between the advection term in \prettyref{eq:energy} and the membrane energy in the F\"oppl-von K\'arm\'an model; likewise the enstrophy term from \prettyref{eq:energy} is to be compared with the bending energy there.
Such analogies are useful routes for the transfer of mathematical methods and theoretical techniques, and we imagine that other such connections are waiting to be found.

\smallskip
\noindent
{\it Acknowledgements --}
We thank R.~V. Kohn and A.~N. Souza for helpful discussions. This work was supported by NSF Awards DGE-0813964 (IT) and DMS-1515161 (CRD), a Van Loo Postdoctoral Fellowship (IT) and a Guggenheim Foundation Fellowship (CRD).

\bibliographystyle{apsrev4-1}
\bibliography{wall2wallrefs}

\end{document}